\documentclass[%
 reprint,
 amsmath,amssymb,
 aps, pre
]{revtex4-2}
\usepackage{caption}
\usepackage{subcaption}
\usepackage{graphicx}
\usepackage{dcolumn}
\usepackage{multirow}
\usepackage{comment}
\usepackage{booktabs}
\usepackage{bm}
\usepackage{tikz}
\usepackage{pgfplots}
\usetikzlibrary{plotmarks}
\pgfplotsset{compat=1.18}
\usepackage{stackengine}
\begin{document}
\preprint{APS/123-QED}
\title{Decoherence of Morse Oscillator in the Presence of Dissipationless Environment}
\author{Titir Mukherjee}
\email{tm21rs050@iiserkol.ac.in}
\affiliation{Department of Physical Sciences, Indian Institute of Science Education and Research Kolkata, Nadia 741246, India}
\author{Arnab Acharya}%
\email{dr.arnab.acharya@gmail.com}
\affiliation{Visvesvaraya National Institute of Technology, Nagpur 440010, India}%

\author{Sushanta Dattagupta}
\email{sushantad@gmail.com}
\affiliation{Sister Nivedita University, New Town, Kolkata 700156, India}
%

\date{\today}

\begin{abstract}
The much-studied Morse oscillator (MO) is couched here in
the context of an open quantum system, in which the
interaction with the quantum environment, however, is taken
to commute with the subsystem Hamiltonian. The result is
decoherence sans dissipation because of dephasing in the
off-diagonal elements of the reduced density operator. The
analytical results are numerically computed for a range of
parameters, for different attributes of decoherence. Finally,
comparison is made for the corresponding harmonic system,
in order to highlight the significance of anharmonicity in the MO, as far as dependencies on the temperature and the
environmental coupling are concerned.
\end{abstract}


\maketitle
\section{Introduction}
The Morse potential, introduced in 1929 by Philip M. Morse \cite{morse1929, morse1934}, has been a foundational model in theoretical chemistry, providing a simplified yet insightful description of the vibrational dynamics of diatomic molecules \cite{Dahl1988}. Unlike the harmonic oscillator (HO), the Morse potential incorporates anharmonicity, the ability of bonds to break, and the existence of unbound states, making it a more realistic representation of molecular behaviour. Although more complex potentials, like the Morse/Long-range(MLR) potential \cite{leroy2009}, are now commonly used in modern spectroscopy, the Morse oscillator (MO) remains a powerful tool for investigating fundamental questions in quantum mechanics of bonds \cite{Goggin1987, Suzuki1999}.
One such question is the transition from the quantum to the classical world, a puzzle that has intrigued physicists for decades \cite{schlosshauer2007}. At the core of this transition is the phenomenon of decoherence - the loss of quantum coherence, or the ability of a system to exhibit interference between superposed states, due to interactions with an environment \cite{Walker1977, Lima2006}. Traditionally, decoherence has been attributed primarily to irreversible coupling with a dissipative environment, leading to energy loss from the quantum system \cite{Cao2001, Keshavamurthy2000}. However, recent work has challenged this view. 

Ford and O'Connell \cite{Ford2001} demonstrated that at high temperatures, where thermal energy can outweigh dissipative effects, decoherence can occur rapidly even with minimal coupling to the environment \cite{Unruh1995, Gangopadhyay1992}. This non-intuitive concept of decoherence without dissipation has spurred investigations in various systems \cite{Banerjee2007, Karmakar2012, foldi2003}, yet a comprehensive understanding has not yet been reached \cite{Gobert2004, Unruh1975}.

This study aims to examine the dynamics of decoherence without dissipation in the context of the one-dimensional MO potential. This system offers a well-defined "quantum sandbox" with established theoretical frameworks and analytical solutions \cite{Kais1990}, providing a robust platform for scrutinising this unconventional mechanism of decoherence \cite{Ackerhalt1985, Nieto1979}. Moreover, MO, unlike HO, possesses a finite number of bound states and a continuum of unbound states \cite{Merzbacher1998, cohentannoudji1977}, allowing for more complex decoherence behaviour \cite{Sakurai1994, Shankar1994}. The Morse potential is given by,
\begin{equation}
    V(x) = D_e ( 1-e^{-a(x-x_e)} )^2,
\end{equation} 
where $D_e$ represents the Morse potential depth and $a$ represents the anharmonicity. The anharmonic nature of the Morse potential will lead to unique features in its decoherence behaviour, differentiating it from the extensively studied harmonic oscillator. Understanding decoherence in this system could provide crucial insights into the quantum-to-classical transition in molecular systems \cite{Landau1977, Baym1969}, a fundamental problem in quantum mechanics that has implications for our understanding of the nature of reality \cite{Fermi1961, Dirac1958}.

Furthermore, this study could contribute to the development of more accurate models for molecular dynamics, particularly in condensed phases \cite{Griffiths2005, Landau1977}, which could lead to advances in fields such as materials science, drug discovery, and nanotechnology \cite{Reichl1998, Reif1965}. Finally, the design of future quantum technologies that require maintaining coherence for extended periods \cite{Zurek1991, Zeilinger2000}, such as in quantum computers, sensors, and communication networks, depends heavily on a deeper understanding of decoherence and its mechanisms \cite{Schlosshauer2004, Giulini1996}.

A driven Morse oscillator arises when a diatomic molecule interacts with a time-varying field. Such driven models have been used to investigate uncertainty products \cite{Keshavamurthy2000, gangopadhyay2001dissipationless}, dissipative dynamics \cite{Graham1991}, classical chaos \cite{gangopadhyay2001dissipationless, Denisov2021}, coherent states \cite{Sage1978}, and the influence of the environment on wave packet evolution \cite{Thiele1961, Oxtoby1976}. Studying decoherence in the one-dimensional MO potential allows for examining the interplay between anharmonicity and environmental interactions, providing insights into the quantum-to-classical transition and its implications for various research areas.

In section II, we describe in detail the mathematical formulation of decoherence without dissipation \cite{Dattagupta2004}. Before going into that, we need to realise where this can occur in physical systems. A prominent example of dissipationless decoherence occurs in nuclear magnetic resonance (NMR). A large static magnetic field causes the Zeeman splitting of nuclear spin levels, while a perpendicular radio-frequency (RF) field induces transitions between these levels. 
If the external static field is strong enough, the
possible off-diagonal elements in the hyperfine interaction
between the nucleus and its surrounding electrons are
suppressed. In that case, the hyperfine coupling commutes
with the Zeeman term. It is indeed through the hyperfine
coupling that bath-induced fluctuations occur through the
electronic spin. But, because of the commutativity implied
above, we end up with dephasing but no energy dissipation. Along similar lines, we consider here an environmental
coupling that contains the Morse Hamiltonian itself, leading to dissipationless decoherence. In Section III, we introduce the measures used to quantify decoherence. In Section IV, we investigate the rate of decoherence and its dependence on the dissipation strength and temperature. We also report a significant increase in the decoherence time if one uses MO instead of HO.

\section{Decoherence without Dissipation}

A closed quantum system remains coherent over time. However, it loses its coherence when it interacts with the environment. Usually, we study decoherence as an effect of dissipative interactions with the environment, where the interaction Hamiltonian is off-diagonal in the product Hilbert space of the subsystem and the heat bath, even though the subsystem and heat bath are individually diagonal. Due to this interaction, energy transfer occurs between the heat bath and the subsystem. However, the bath remains in a thermal equilibrium state due to its large size \cite{breuer2002theory}.

In this paper, we present an alternative scenario of decoherence, wherein the interaction Hamiltonian is diagonal in the subsystem representation but remains off-diagonal in the bath representation. Specifically, we consider a case where $[H, H_s] = 0$, $[H_s, H_B] = 0$, but $[H_I, H_B] \neq 0$, where $H$, $H_S$, $H_I$, and $H_B$ are the total, subsystem, interaction, and bath Hamiltonians, respectively. A specific example of this is where the interaction Hamiltonian is proportional to the system Hamiltonian itself \cite{gangopadhyay2001dissipationless, Dattagupta2004}. The total Hamiltonian can be written as follows,
\begin{eqnarray}
\centering
&H = H_s+ H_B+H_I  \\ \nonumber
&= H _{S}+\Sigma_{j}\left\{ p _{j}^2 / 2 m_{j}+1 / 2 m_j \omega_j^2\left( x _{j}+C_{j} H _{s} / m_j \omega_j^2\right)^2\right\},
\end{eqnarray}
where, following usual practice \cite{caldeira1983path,ford1988quantum}, we take the heat bath as a large collection of non-interacting harmonic oscillators, $p_j$, $m_j$, and $\omega_j$ are the momentum, mass, and natural frequency of the $j$-th mode of the environment, respectively. The last term is added to ensure translational invariance of the total system. However, this term makes no meaningful contribution as it disappears under a unitary transformation, as shown below.
Now, the density operator evolves according to the Liouville equation:
\begin{equation}
\rho(t) = e^{-i H t} \rho(0) e^{i H t}
.\end{equation}
As usual, we assume that the initial state of the composite system is separable, i.e.,
\begin{equation}
\rho(0) = \rho_B \otimes \rho_s(0)
.\end{equation}
The density operator for the heat bath is a thermal state (Gibbs density matrix), given by,
\begin{equation}
\rho_B = \frac{e^{-\beta H_B}}{\text{Tr}_B e^{-\beta H_B}},
\end{equation}
where, $\beta=(kT)^{-1}$ represnts the temperature. Let us now define a unitary transformation that generates translation in the bath:
\begin{equation}{\label{trans}}
S=exp\left[ H_s\sum_k g_k (b_k-b_k^\dagger)\right]
.\end{equation}
This transforms the Hamiltonian as follows,
\begin{equation}
\Tilde{H}=SHS^{-1}=H_s+\sum_k \omega_k b_k(t) b_k^\dagger(t)
.\end{equation}
Using this transformation, the density operator $\rho(t)$ evolves according to,
\begin{equation}
\rho(t) = S^{-1}e^{-i\Tilde{H} t}S \left( \rho_B \otimes \rho_s(0) \right) S^{-1}e^{i \Tilde{H} t}S
.\end{equation}
Taking the partial trace over the bath degrees of freedom, the reduced dynamics of the system becomes,
\begin{equation}
\rho_s=Tr_B(\rho(t)) = Tr_B(S^{-1}e^{-i\Tilde{H} t}S \left( \rho_B \otimes \rho_s(0) \right) S^{-1}e^{i \Tilde{H} t}S)
.\end{equation}

We now find the time evolution of the $(n,m)$-th element of the reduced density operator for the system, $\rho_{s, nm}(t)$, in the eigenbasis of $H_S$,
\small
\begin{eqnarray}
\rho_{s, n m}(t)&=&e^{-i\left(E_n-E_m\right) t} \\&\times&\operatorname{Tr}_B\left(\rho_B S _m^{-1} e^{i H _B t} S _m S _n^{-1} e^{-i H _B t} S _n\right) \rho_{s, n m}(0) \nonumber.
\label{eqn: rho}
\end{eqnarray}
\normalsize
where,
\begin{equation}
S _n=\exp \left[E_n \sum_k \frac{g_k}{\omega_k}\left(b_k^{\dagger}(t)-b_k(t)\right)\right],
\end{equation}
\begin{equation}
H_s |n\rangle = E_n |n\rangle
.\end{equation}

After expanding $S_n$, we arrive at the following expression,
\begin{eqnarray}
e^{i H _B t} S _m S _n^{-1} e^{-i H _B t}= \nonumber \\ 
\exp \left\{\left(E_m-E_n\right) \sum_k \frac{g_k}{\omega_k}\left[b_k^{\dagger}(t)-b_{ k }(t)\right]\right\}.
\end{eqnarray}
To proceed, we require the time evolution of the bath operators. Using Heisenberg's equation of motion and considering the bath as a large collection of harmonic oscillators, we get the following equation,
\begin{equation}
\begin{aligned}
b_k^{\dagger}(t) & =e^{i \omega_k t} b_k^{\dagger}, \\
b_k(t) & =e^{-i \omega_k t} b_k .
\end{aligned}
.\end{equation}

All the $S$ operators are exponential operators whose exponents are linear combinations of $b_k$ and $b_k^\dagger$. For two operators $\hat{A}$ and $\hat{B}$, if $[\hat{A},\hat{B}] = \text{(c-number)}$, then we can use the identity:
\begin{equation}
  e^{(\hat{A}+\hat{B})}=e^{-\frac{[\hat{A}, \hat{B}]}{2}} e^{\hat{A}} e^{\hat{B}} 
.\end{equation}
Using this identity for $b_k$ and $b_k^\dagger$, one can simplify the bath part of the equation~\ref{eqn: rho} as following,
\small
\begin{eqnarray}
&& S _m^{-1} e^{i H _B t} S _m S _n^{-1} e^{-i H _B t} S _n \nonumber\\& = &
\exp \left\{-i\left(E_m^2-E_n^2\right) \sum_k\left(\frac{g_k}{\omega_k}\right)^2 \sin \left(\omega_k t\right)\right\} \\
& \times& \exp \left\{\left(E_m-E_n\right) \sum_k \frac{g_k}{\omega_k}\left[\left(e^{i \omega_k t}-1\right) b_k^{+}+\left(1-e^{-i \omega_k t}\right) b_k\right]\right\} \nonumber.
\end{eqnarray}
\normalsize
By following the standard procedure described in \cite{Dattagupta2004,gangopadhyay2001dissipationless}, we finally arrive at the following equation,
\small
\begin{eqnarray}
    \operatorname{Tr}_B\left[\rho_{B} e^{\left(x_k b_{ k }-x_k^* b_k^{+}\right)}\right]=\exp \left[\frac{\left|x_{ k }\right|^2}{2} \operatorname{coth}\left(\frac{\beta \omega_{ k }}{2}\right)\right].
\end{eqnarray}
\normalsize
Thus, the final expression becomes
\small
\begin{eqnarray}
\rho_{s, n m}(t)&=& \left.\exp \left[-i\left(E_n\right.-{E_m}\right) t\right] \exp \left[i\left(E_n^2-{E_m^2}\right) \eta(t)\right] \nonumber \\&\times& \exp \left[-\left(E_n-E_m\right)^2 \gamma(t)\right]{\rho_{s, n m}}(0).
\end{eqnarray}
\normalsize
where
\begin{equation}
    \begin{gathered}
\eta(t)=-\sum_k\left(\frac{g_k}{\omega_k}\right)^2 \sin \left(\omega_k t\right), \\
\gamma(t)=\sum_k\left(\frac{g_k}{\omega_k}\right)^2 \sin ^2\left(\frac{\omega_k t}{2}\right) \operatorname{coth}\left(\frac{\beta \omega_k}{2}\right).
\end{gathered}
\end{equation}

To proceed with numerical calculations, we must evaluate $\eta(t)$ and $\gamma(t)$. Since the bath consists of a large number of modes, we consider a continuous frequency distribution. The summations over $k$ in $\eta(t)$ and $\gamma(t)$ then convert into integrals
\begin{equation}
\sum_k g_k^2 f\left(\omega_k\right) \rightarrow \int_0^{\infty} d \omega J(\omega) f(\omega)
.\end{equation}
We choose the standard Ohmic spectral density,
\begin{equation}
   J(\omega)=\Gamma \omega e^{-\omega / \omega_c}
.\end{equation}
where $\Gamma$ is the coupling strength and $\omega_c$ is the cutoff frequency.

The expressions for $\eta(t)$ and $\gamma(t)$ then become,
\begin{equation}
\begin{gathered}
\eta(t)=-\int_0^{\infty}\frac{d\omega J(\omega)\sin \left(\omega t\right)}{\omega^2}, \\
\gamma(t)= \int_0^{\infty}\frac{d\omega J(\omega)\sin ^2\left(\frac{\omega t}{2}\right)\operatorname{coth}\left(\frac{\beta \omega}{2}\right)}{\omega^2}.
\end{gathered}
.\end{equation}
Once we compute $\eta(t)$ and $\gamma(t)$ numerically, we can obtain $\rho(t)$ at any time in the eigenbasis of $H_S$ and study the system’s decoherence.
\begin{figure}[h!]
     \centering
\includegraphics[width=0.35\textwidth]{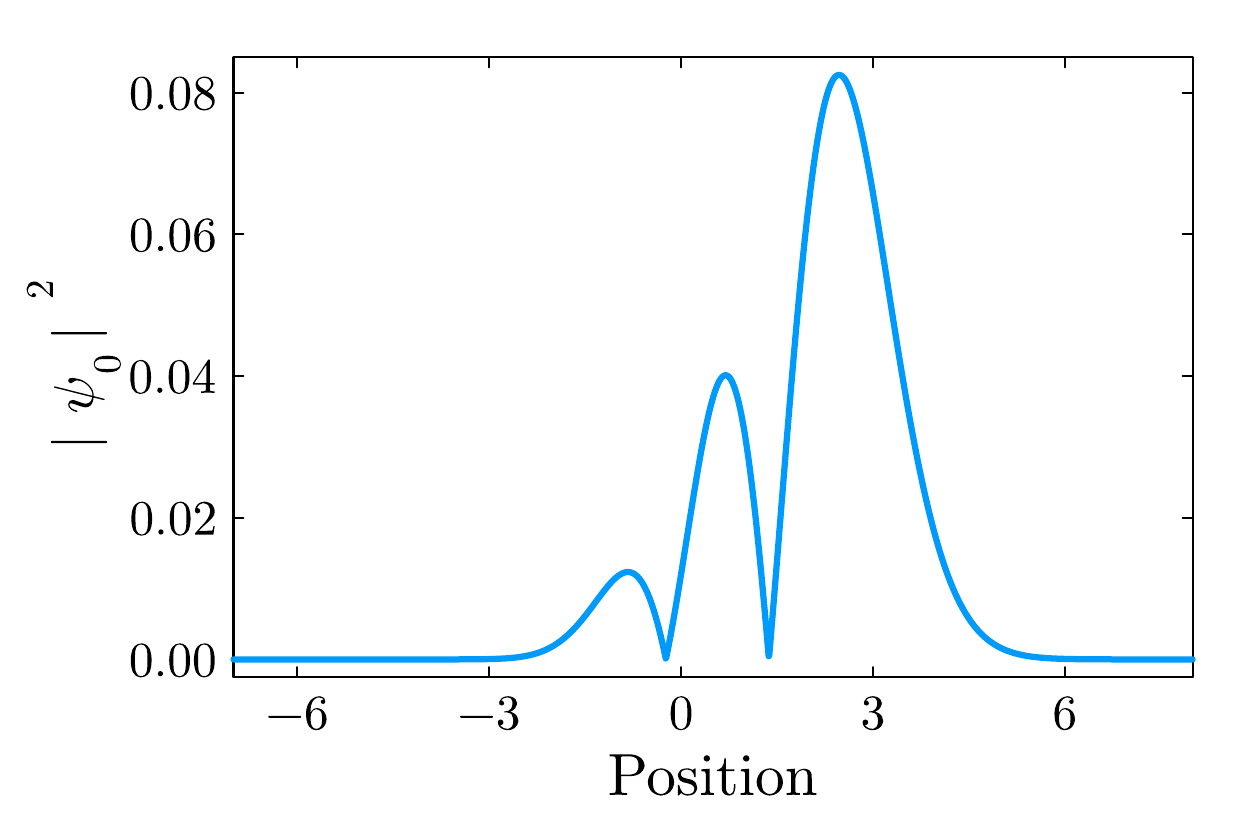}
\caption{Initial state of the system, which is the coherent state of the Morse oscillator with $\mu=2.4$, $\sigma=0.5$}
\label{fig: initial}
\end{figure}
\section{Results}
\subsection{Measures uses}
In this paper, we study decoherence without dissipation for a Morse oscillator. For that purpose, we choose the Gaussian coherent state of the Morse oscillator (see Fig.~\ref{fig: initial}) \cite{morse1990coherent,morse2008coherent} as the initial state, which is given by 
\begin{eqnarray}
|\mu, \sigma\rangle = \frac{1}{\sqrt{\mathcal{N}(\mu, \sigma)}} \sum_{n \in I} e^{-\frac{(n - \mu)^2}{4\sigma^2}} |\psi_n\rangle, \quad\\
\text{where,} \hspace{0.5cm}
\mathcal{N}(\mu, \sigma) = \sum_{n \in I} e^{-\frac{(n - \mu)^2}{2\sigma^2}}. \nonumber
\end{eqnarray}

To understand how a dissipationless environment causes decoherence in the system, we consider a few standard measures.

The first measure we use is the temporal behaviour of the survival probability \cite{gangopadhyay2001dissipationless}, defined as,
\begin{equation}
    P(t) = \operatorname{Tr}[\rho(0)\rho(t)]
.\end{equation}

Another popular measure is the temporal behaviour of purity \cite{decohernce2018measure}.
\begin{equation}
    D(t) = \operatorname{Tr}(\rho(t)^2)
.\end{equation}
In literature, Von Neumann entropy $S(\rho) = -\operatorname{Tr}(\rho \log \rho)$ is a measure of how mixed or uncertain a quantum state is, and it increases as the system decoheres.
There are two other useful coherence measures discussed in the literature \cite{decohernce2018measure}. The first one is the relative entropy of coherence, defined as follows
\begin{gather}
    C_{e}(\rho) = S\left(\rho{(t)}\right) - S(\rho(0))
    \\=\operatorname{Tr}(\rho(t) \log \rho(t))-\operatorname{Tr}(\rho(0) \log \rho(0))
.\end{gather}
Where the second one is the 2-norm of coherence, defined by the following equation,
\begin{gather}
C_2(\rho) = D(\rho(0)) - D\left(\rho{(t)}\right)\\
=\operatorname{Tr}(\rho(0)^2)-\operatorname{Tr}(\rho(t)^2)
.\end{gather}

We also compute the decoherence time using the definition,
\begin{eqnarray}
\tau_{\text{mn}} &=& \int_0^{\infty} \frac{|\rho_{mn}(t)|}{|\rho_{mn}(0)|} \ dt,
\end{eqnarray}
where $\rho_{mn}$ is the m-th, n-th element of the density matrix. For the element-wise coherence time, we take the smallest among all $\tau_{\text{mn}}$, as suggested in \cite{Fedichkin2003measures}, and define it as $\tau_{\text{element}}$.

\begin{figure}[htbp!]
\begin{subfigure}[b]{0.35\textwidth}
         \centering
         \includegraphics[width=\textwidth]{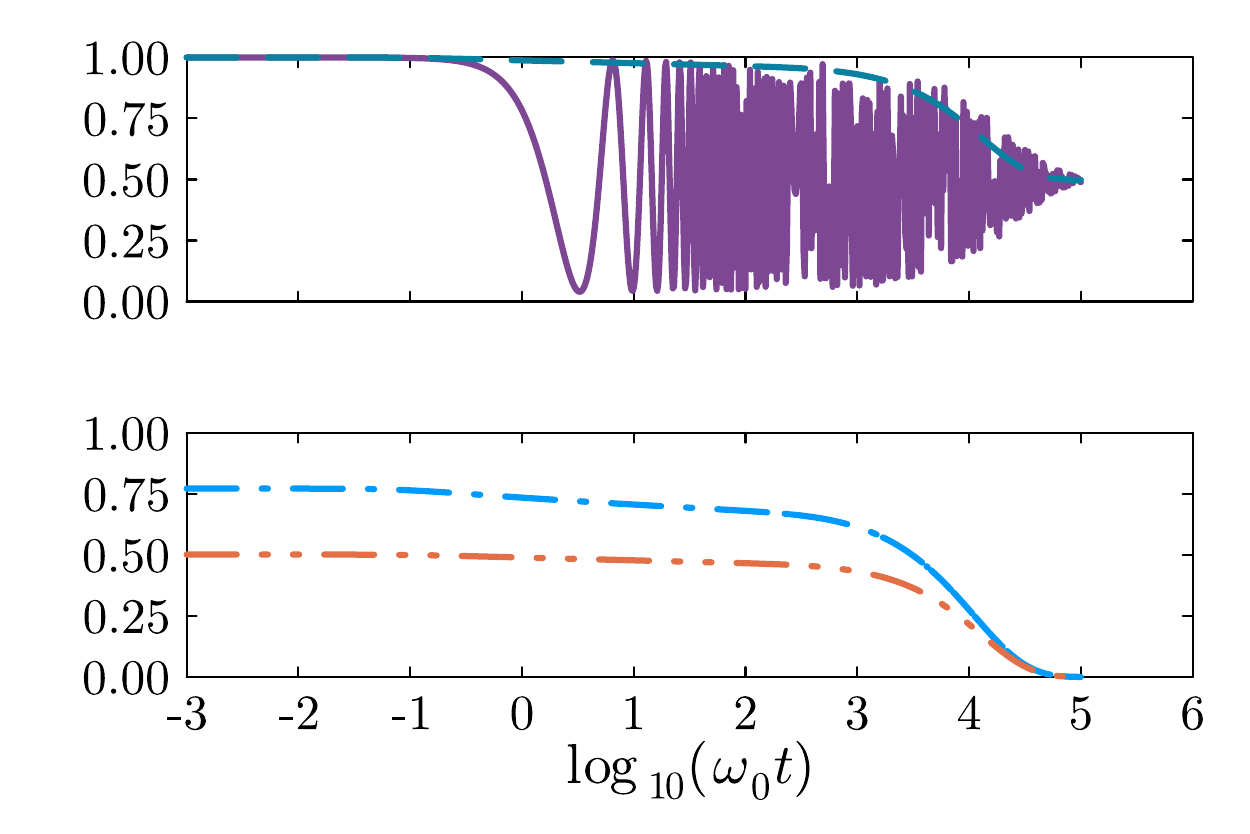}
         \caption{}
     \end{subfigure}
     \centering
     \begin{subfigure}[b]{0.35\textwidth}
         \centering
         \includegraphics[width=\textwidth]{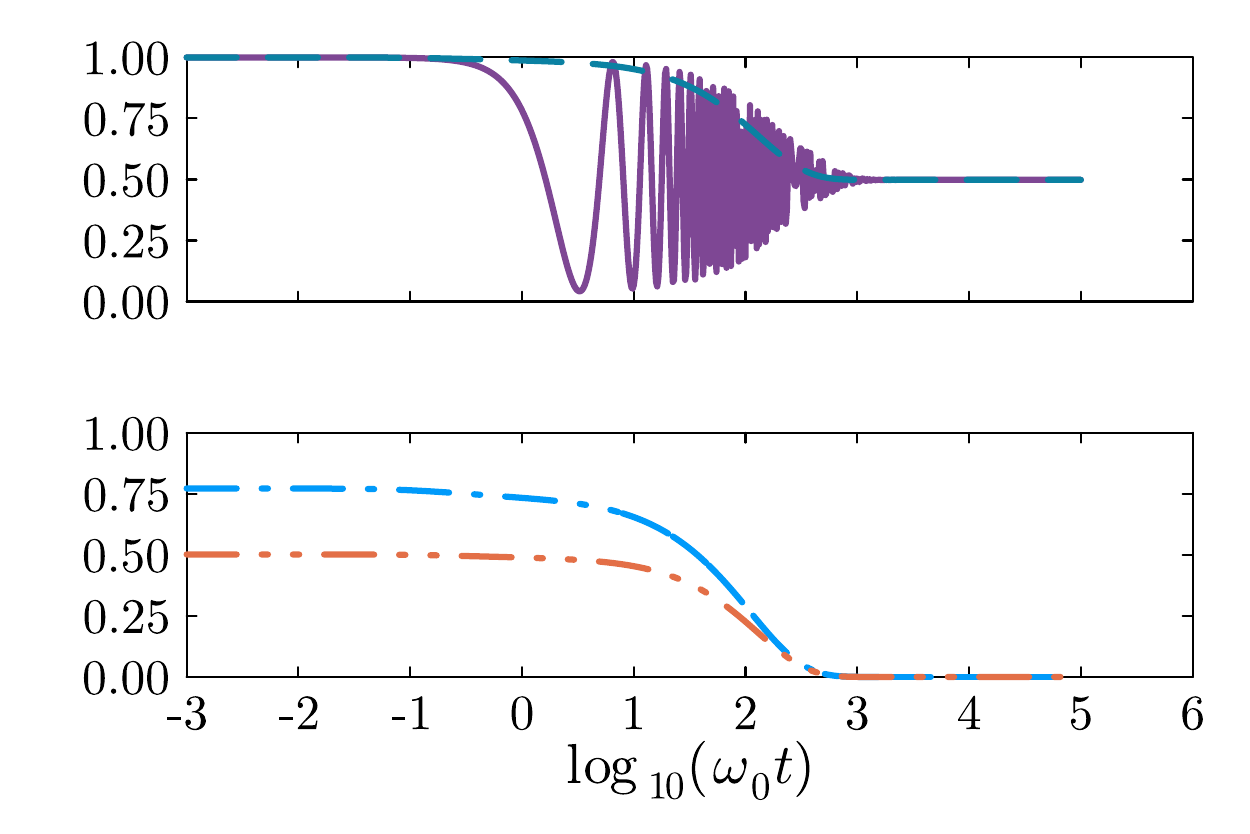}
         \caption{}
     \end{subfigure}
     \begin{subfigure}[b]{0.35\textwidth}
         \centering
         \includegraphics[width=\textwidth]{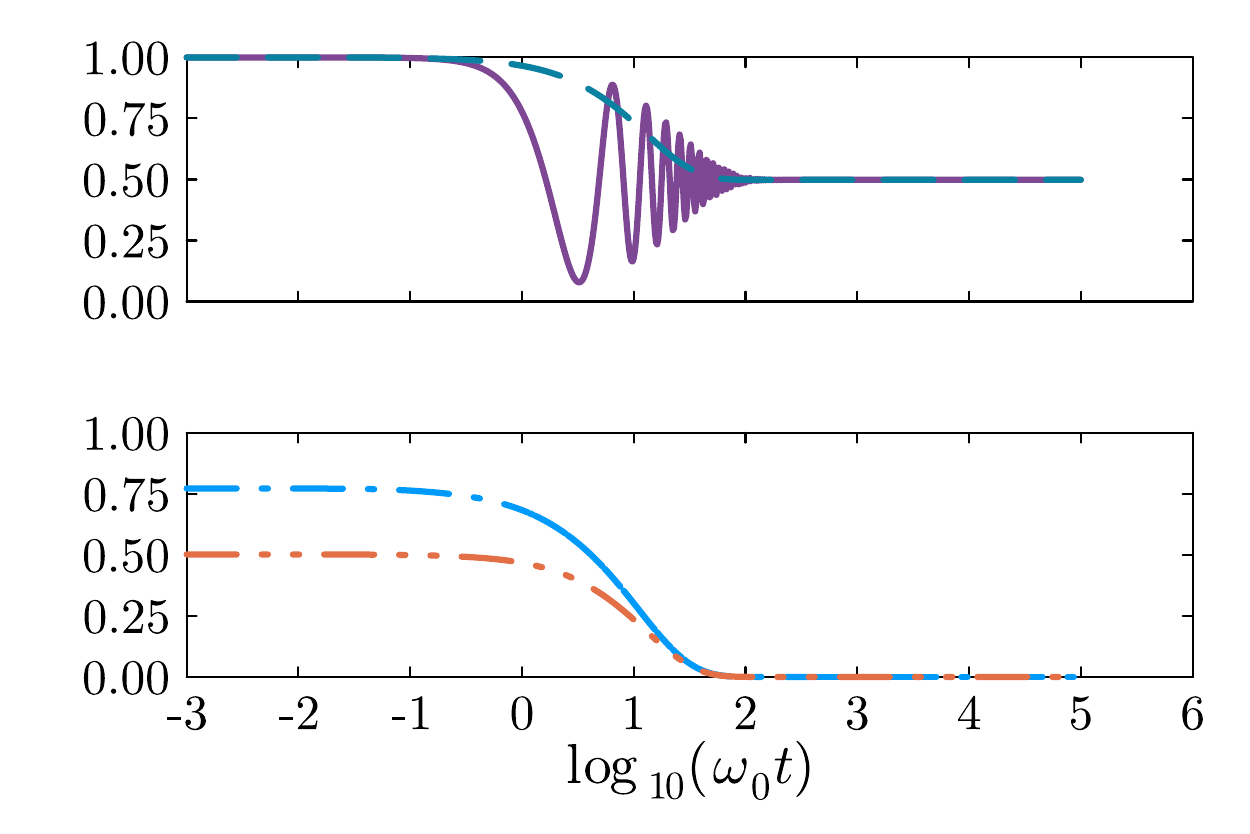}
         \caption{}
     \end{subfigure}
     \begin{subfigure}[b]{0.35\textwidth}
         \centering
         \stackinset{r}{0.001cm}{b}{1.1cm}{\includegraphics[width=0.5\textwidth]{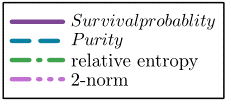}}{\includegraphics[width=\textwidth]{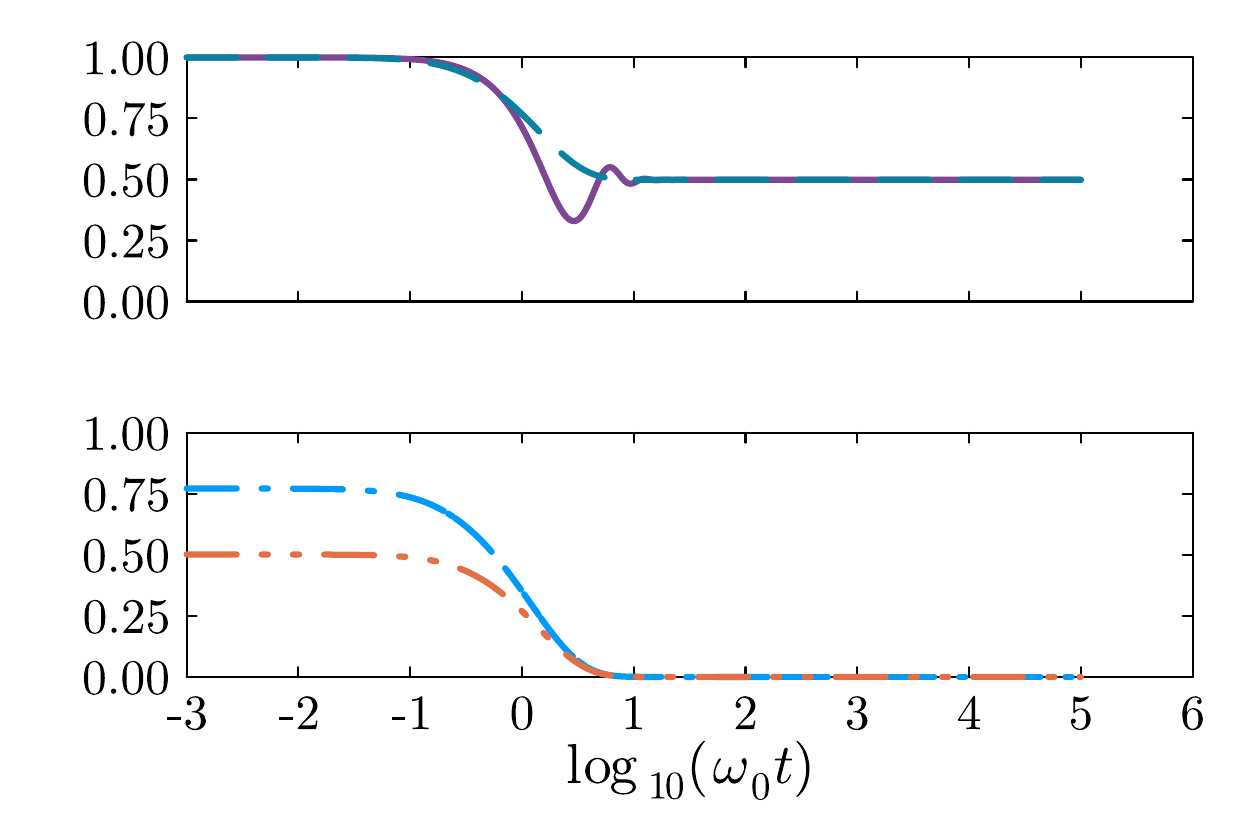}}
         \caption{}
     \end{subfigure}
\caption{Temporal dependence of survival probability (solid line) and purity (dashed line) in the top panel, and relative entropy (dash-dotted line) and 2-norm of decoherence (dash-dot-dotted line) in the bottom panel for the Morse potential. Subfigures show results for different temperatures: (a)~$kT=0.01$, (b)~$kT=1.0$, (c)~$kT=10.0$, and (d)~$kT=100.0$.}
    \label{fig: temp-norm}
\end{figure}
\begin{figure}[htbp!]
\begin{subfigure}[b]{0.35\textwidth}
         \centering
         \includegraphics[width=\textwidth]{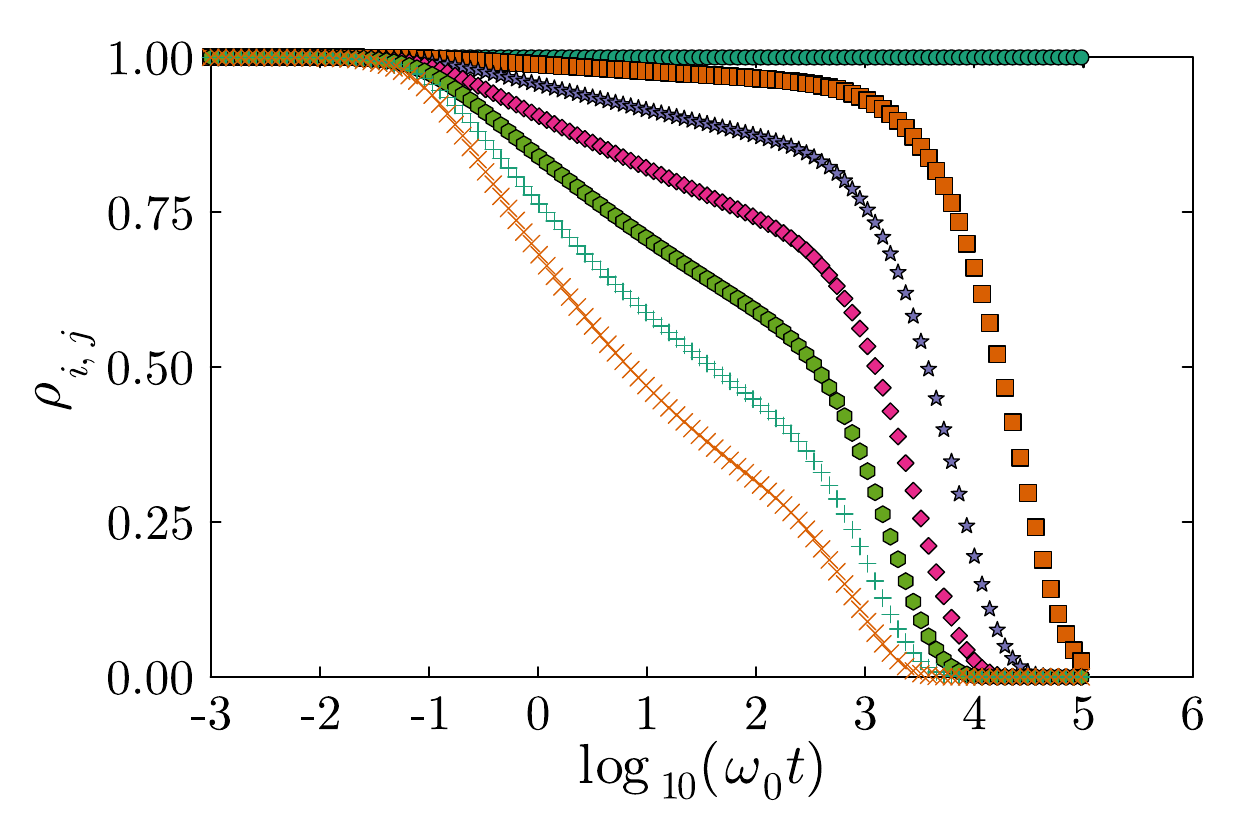}
         \caption{}
     \end{subfigure}
     \centering
     \begin{subfigure}[b]{0.35\textwidth}
         \centering
         \includegraphics[width=\textwidth]{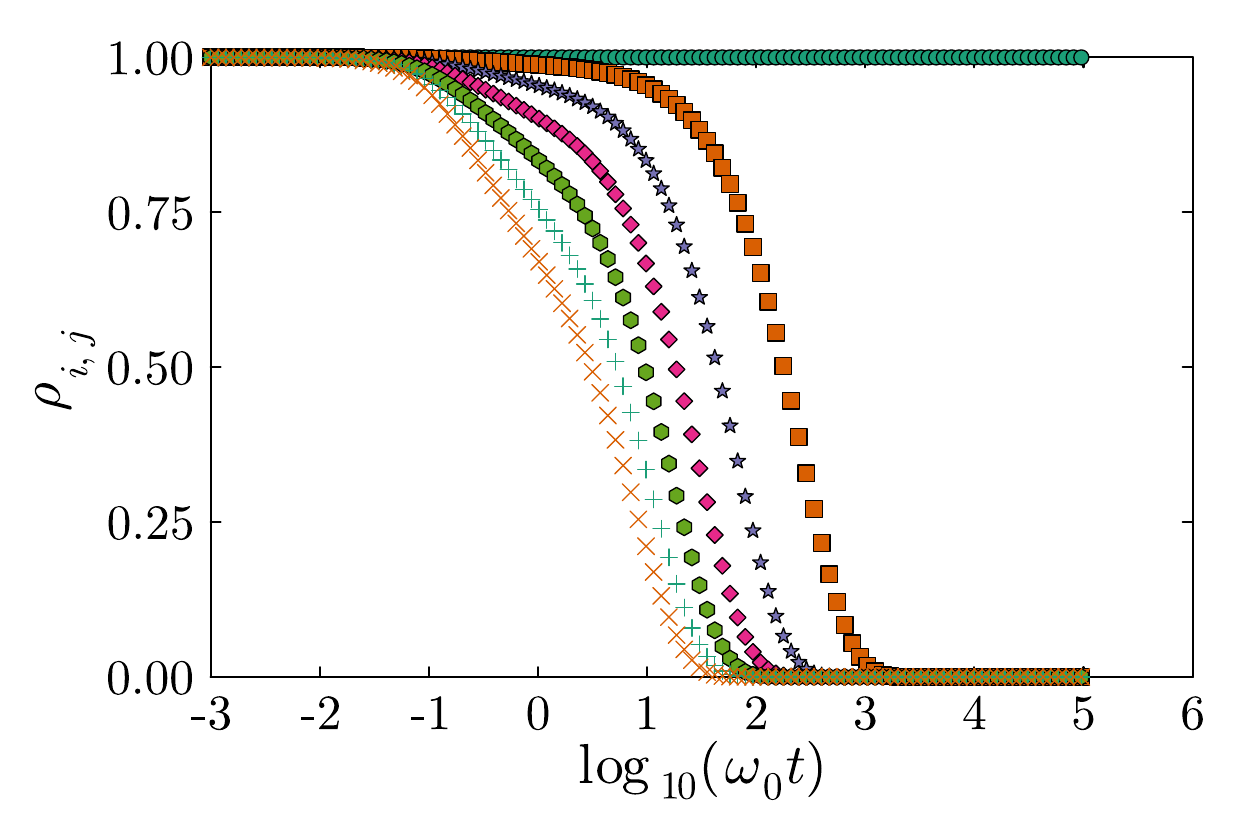}
         \caption{}
     \end{subfigure}
     \begin{subfigure}[b]{0.35\textwidth}
         \centering
         \includegraphics[width=\textwidth]{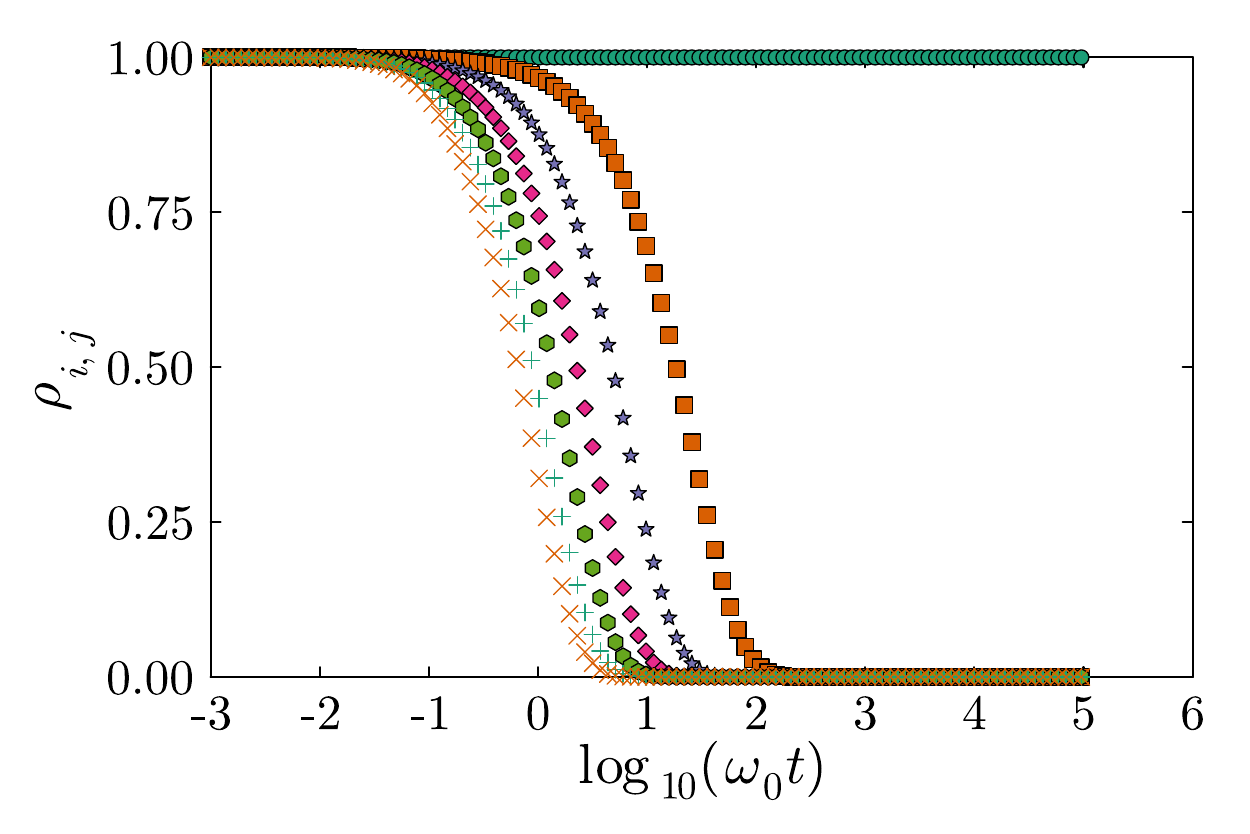}
         \caption{}
     \end{subfigure}
     \begin{subfigure}[b]{0.35\textwidth}
         \centering
          \stackinset{r}{0.001cm}{b}{0.5cm}{\includegraphics[width=0.2\textwidth]{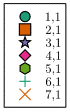}}
         {\includegraphics[width=\textwidth]{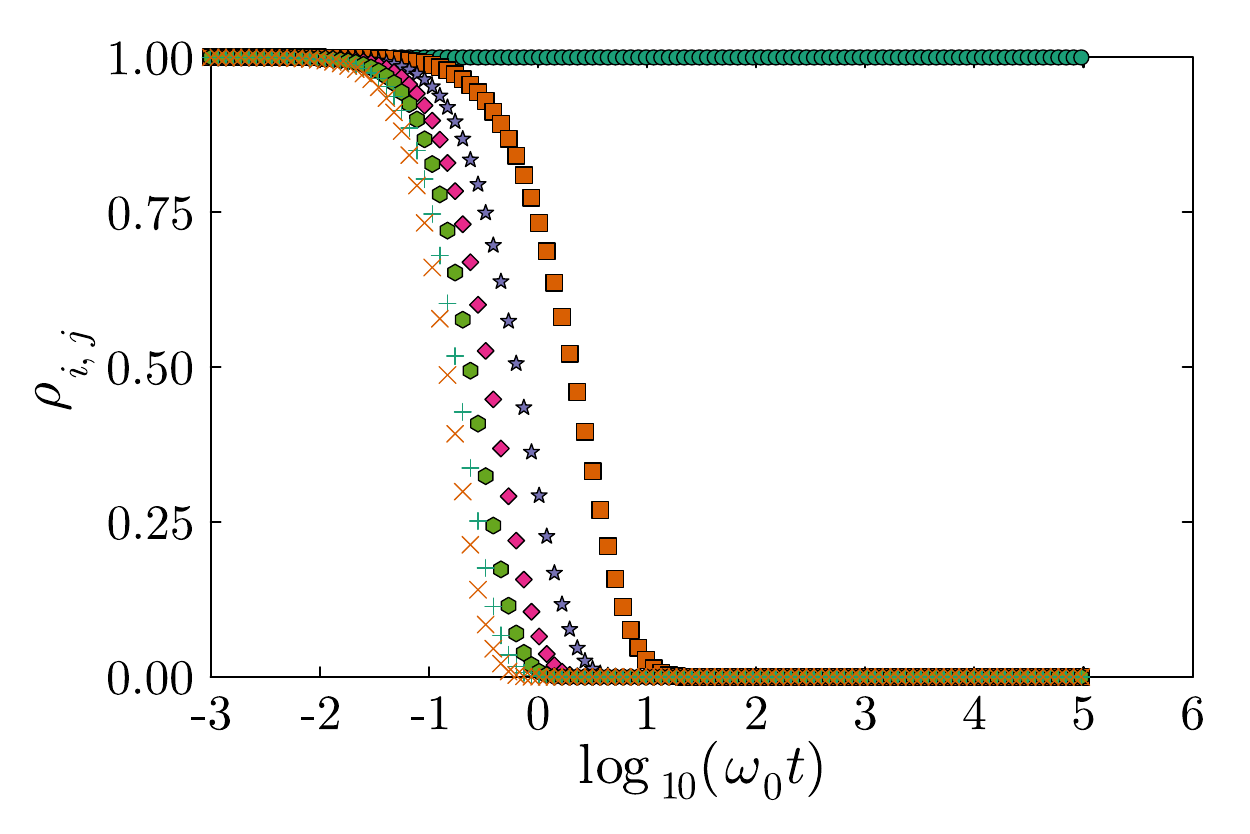}}
         \caption{}
     \end{subfigure}
\caption{Time decay of density matrix elements: $\rho_{1,1}(t)$ (circle), $\rho_{2,1}(t)$ (rectangle), $\rho_{3,1}(t)$ (five-point star), $\rho_{4,1}(t)$ (diamond), $\rho_{5,1}(t)$ (hexagon), $\rho_{6,1}(t)$ (cross), and $\rho_{7,1}(t)$ (x-cross). Sub-figures show results for different temperatures: (a)~$kT = 0.01$, (b)~$kT = 1.0$, (c)~$kT = 10.0$, and (d)~$kT = 100.0$.}
\label{fig: temp-elem}
\end{figure}
\subsection{Numerical simulation results}
The parameters related to the Morse potential—$x_e$ (equilibrium distance), $D_e$ (potential depth), and $a$ (width of the potential)—are taken as $x_e = 1.0$, $D_e = 40.0$, and $a = 0.11$. Hence, $n_{\text{max}}$, the number of maximum accessible eigenstates, and $\omega_0$, the natural frequency, turn out to be $79$ and $1$, respectively. We also choose $\hbar = 1$ and $m = 1$ throughout the simulation. For the initial coherent state, the mean is $\mu = 2.4$ and the standard deviation is $\sigma = 0.5$. The parameters related to the bath are $\omega_c$ (cut-off frequency for bath spectral density), $\Gamma$ (coupling strength for bath spectral density) and $kT$ (temperature).We take $\omega_c = 10 \times \omega_0$. \\
In this paper, we vary $\Gamma$ and $kT$ one by one and show how decoherence sets in the system. For that, we calculate survival probability $P(t)$, purity $D(t)$, the entropy of decoherence $C_e(\rho)$, 2-norm of coherence $C_2(\rho)$, and showed how they changed with time when the system evolved in the presence of dissipationless decoherence.
In Fig.~\ref{fig: temp-norm}, top panel shows the temporal behavior of $P(t)$ and $D(t)$, while bottom panel displays the same for $C_e(\rho)$ and $C_2(\rho)$. From Fig.~\ref{fig: temp-norm}(a) to Fig.~\ref{fig: temp-norm}(d), we gradually increase the temperature ($kT$). Specifically, for Fig.~\ref{fig: temp-norm}(a), $kT = 0.01$; in Fig.~\ref{fig: temp-norm}(b), $kT = 1.0$; in Fig.~\ref{fig: temp-norm}(c), $kT = 10.0$; and in Fig.~\ref{fig: temp-norm}(d), $kT = 100.0$.
From the graphs, it is clear that at higher temperatures, the system loses coherence more rapidly. This suggests that as $kT$ increases, the effect of the environment on the system becomes more pronounced. Another interesting observation is that the oscillatory behaviour of $P(t)$ decreases as $kT$ increases.
\\
In Fig.~\ref{fig: temp-elem}, we examine the time variation of different terms in the density matrix. While the diagonal term does not decay as expected, the off-diagonal terms decay at different rates. The terms closer to the diagonal decay more slowly compared to the farther ones. As $kT$ increases, the decay time for each term decreases, but the difference in decay times between the various terms also becomes smaller.
\\ 
\begin{figure}[hbtp!]
\begin{subfigure}[b]{0.35\textwidth}
         \centering
         \includegraphics[width=\textwidth]{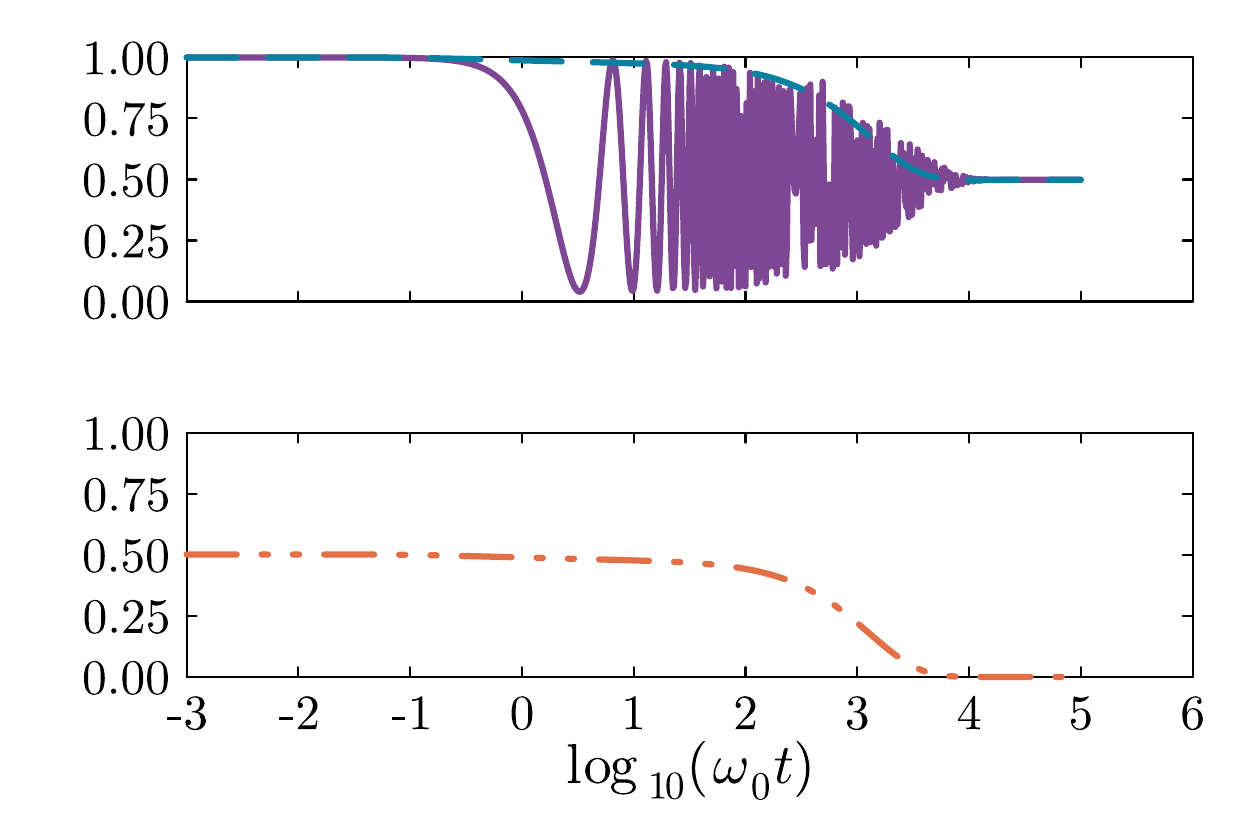}
         \caption{}
     \end{subfigure}
     \centering
     \begin{subfigure}[b]{0.35\textwidth}
         \centering
         \includegraphics[width=\textwidth]{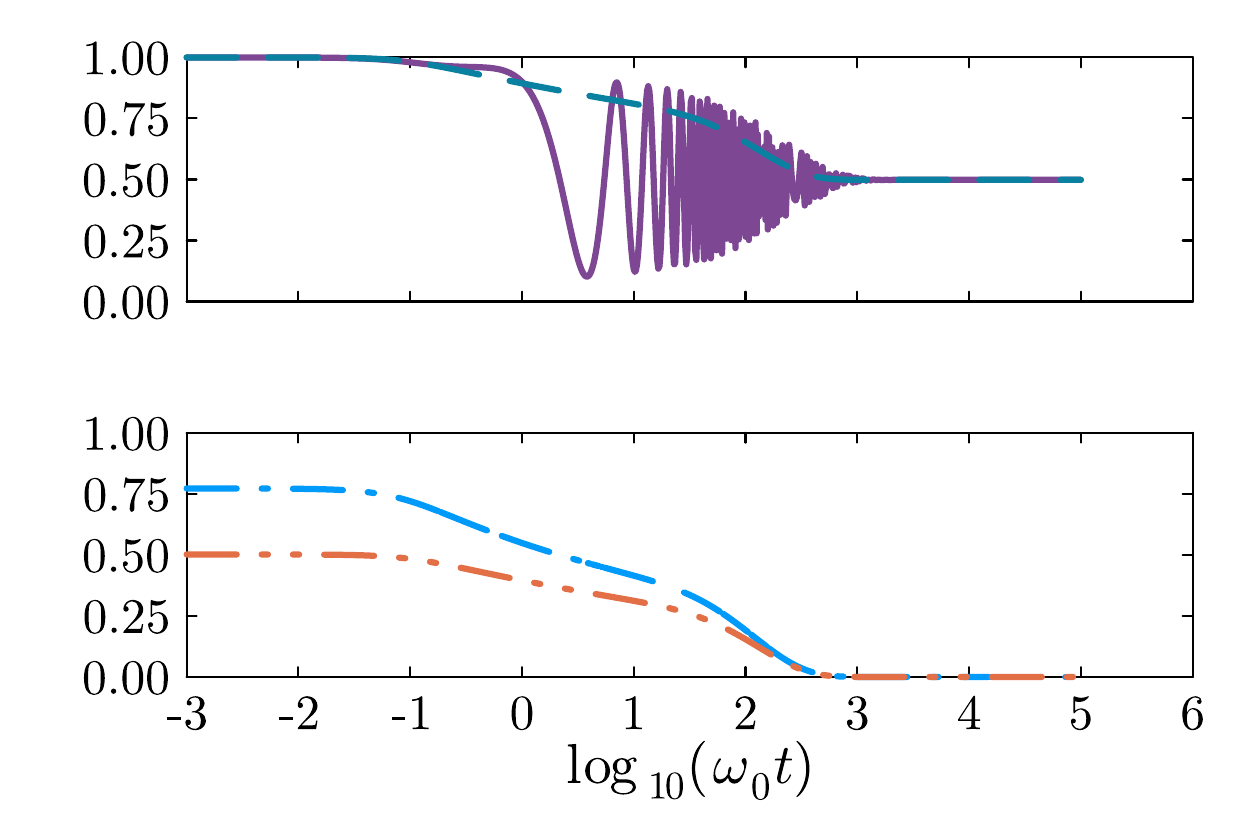}
         \caption{}
     \end{subfigure}
     \begin{subfigure}[b]{0.35\textwidth}
         \centering
         \includegraphics[width=\textwidth]{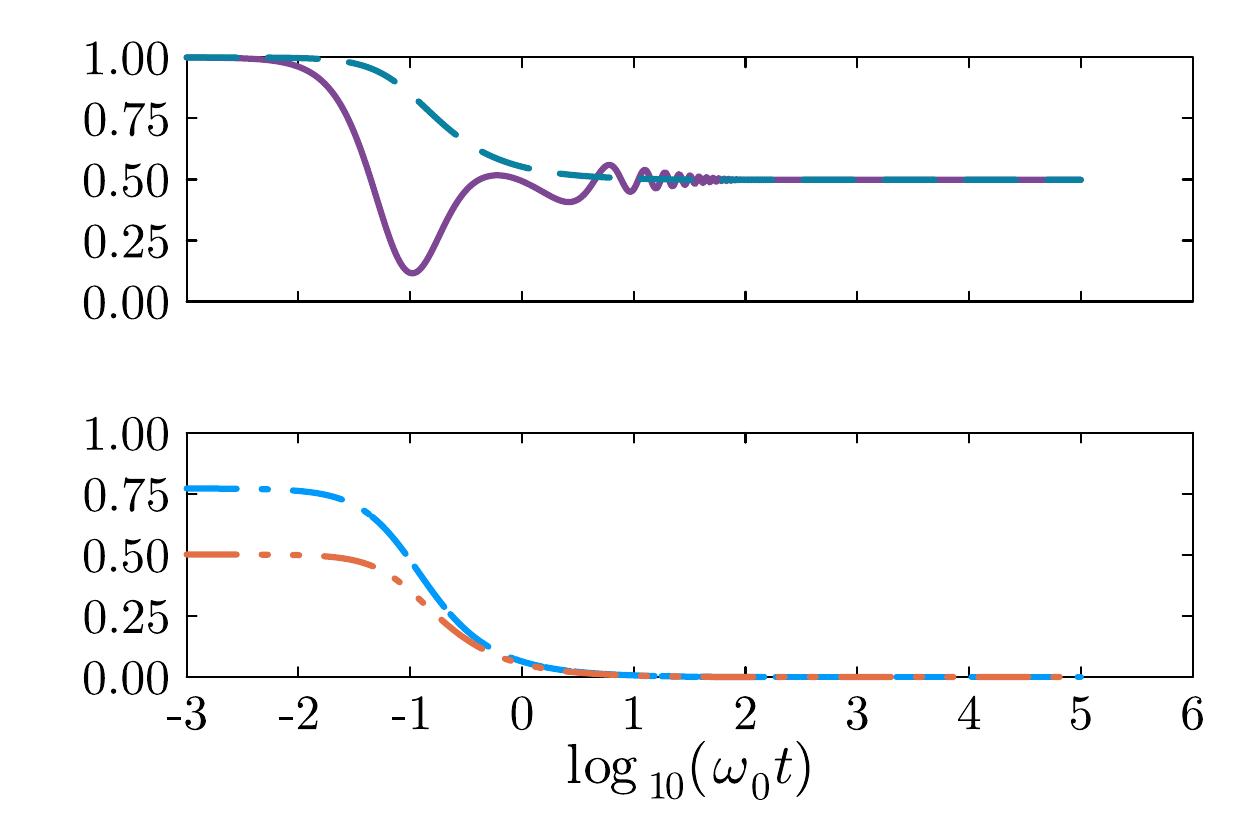}
         \caption{}
     \end{subfigure}
     \begin{subfigure}[b]{0.35\textwidth}
         \centering
         \stackinset{r}{0.001cm}{b}{1.1cm}{\includegraphics[width=0.5\textwidth]{label2.png}}{\includegraphics[width=\textwidth]{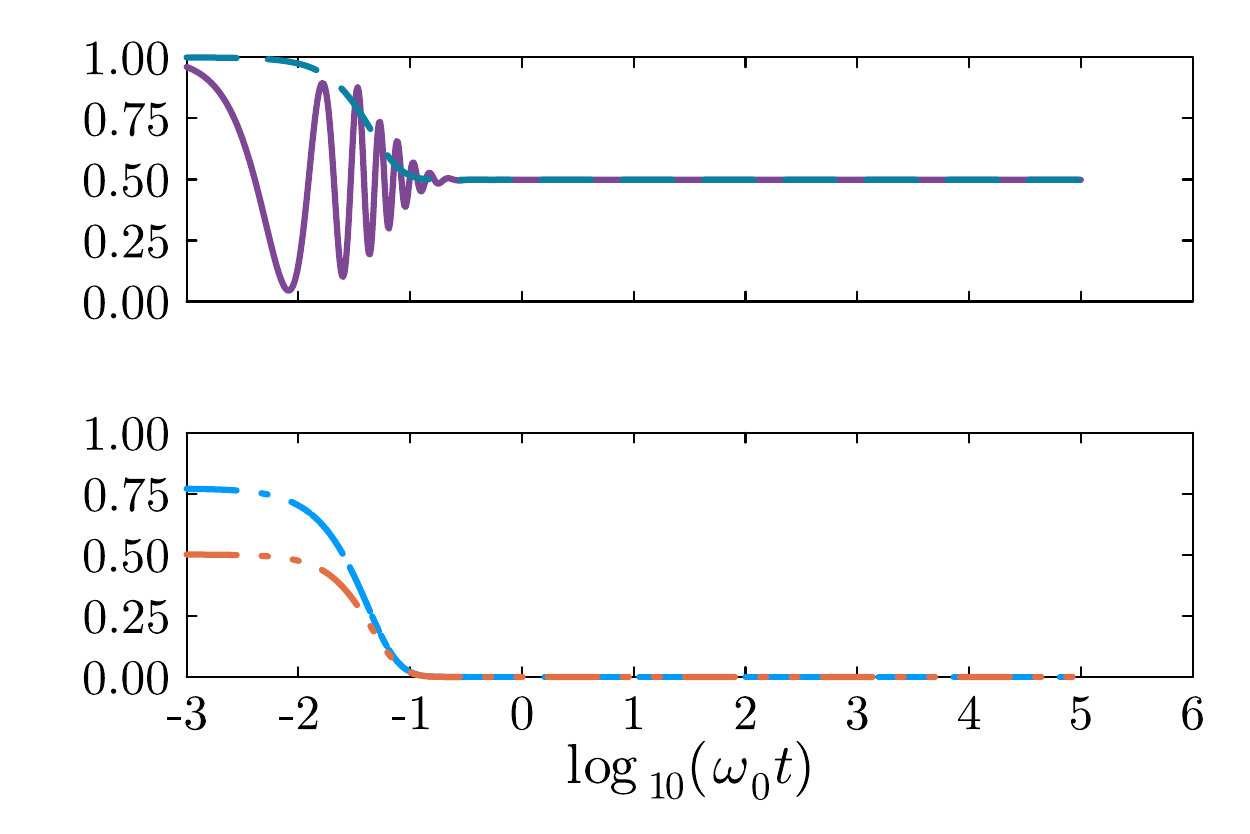}}
         \caption{}
     \end{subfigure}
\caption{Temporal dependence of survival probability (solid line) and purity (dashed line) in the top panel, and relative entropy (dash-dotted line) and 2-norm of decoherence (dash-dot-dotted line) in the bottom panel for the Morse potential. Subfigures show results for different temperatures: (a)~$\Gamma=0.01$, (b)~$\Gamma=0.1$, (c)~$\Gamma=1.0$, and (d)~$\Gamma=10.0$.
}
\label{fig: gamma-norm}
\end{figure}
\begin{figure}[htbp!]
\begin{subfigure}[b]{0.35\textwidth}
         \centering
         \includegraphics[width=\textwidth]{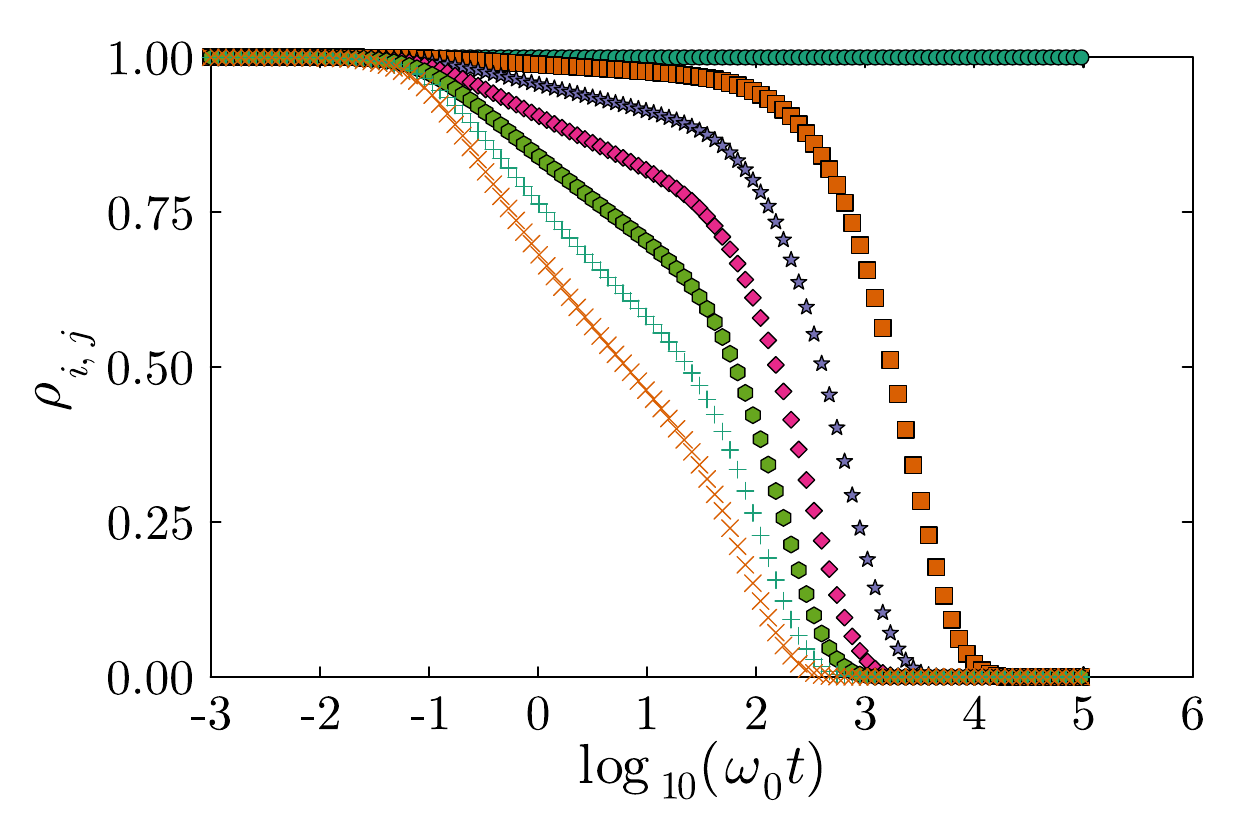}
         \caption{}
     \end{subfigure}
     \centering
     \begin{subfigure}[b]{0.35\textwidth}
         \centering
         \includegraphics[width=\textwidth]{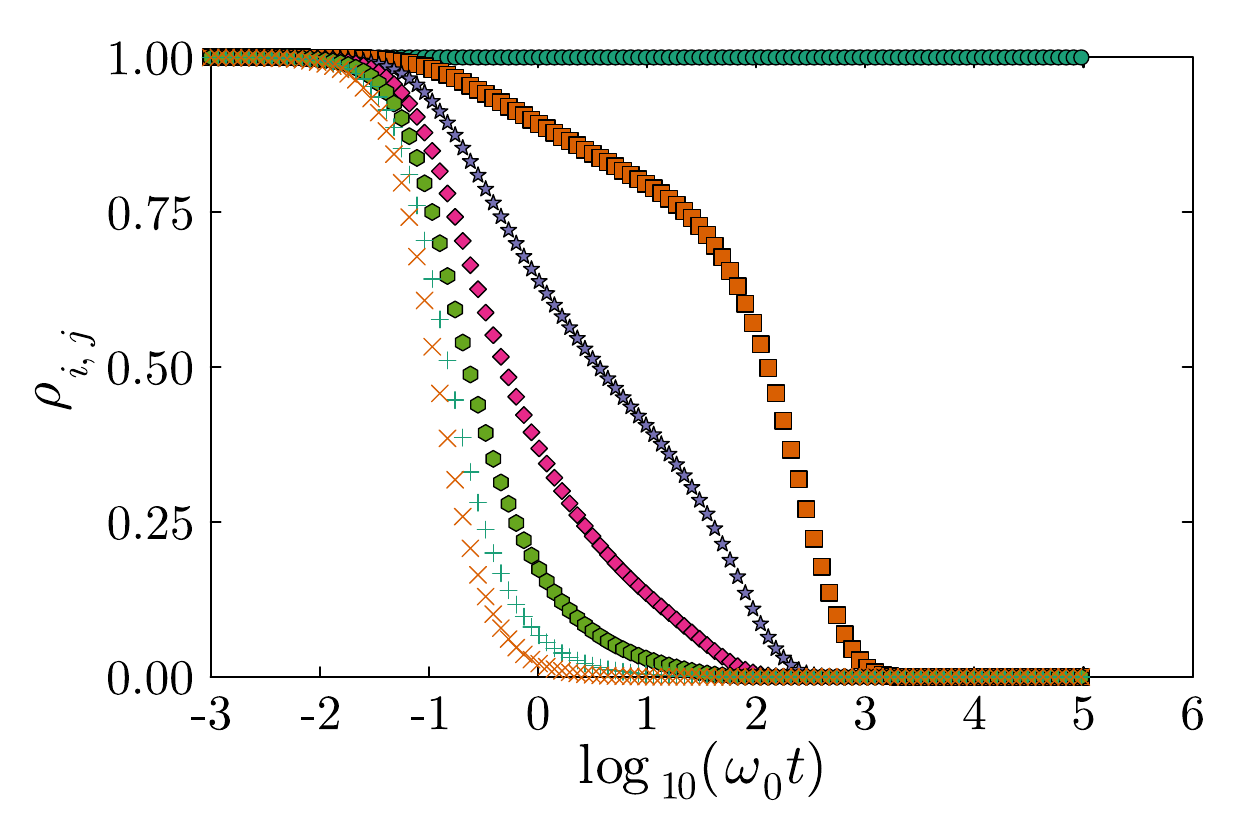}
         \caption{}
     \end{subfigure}
     \begin{subfigure}[b]{0.35\textwidth}
         \centering
         \includegraphics[width=\textwidth]{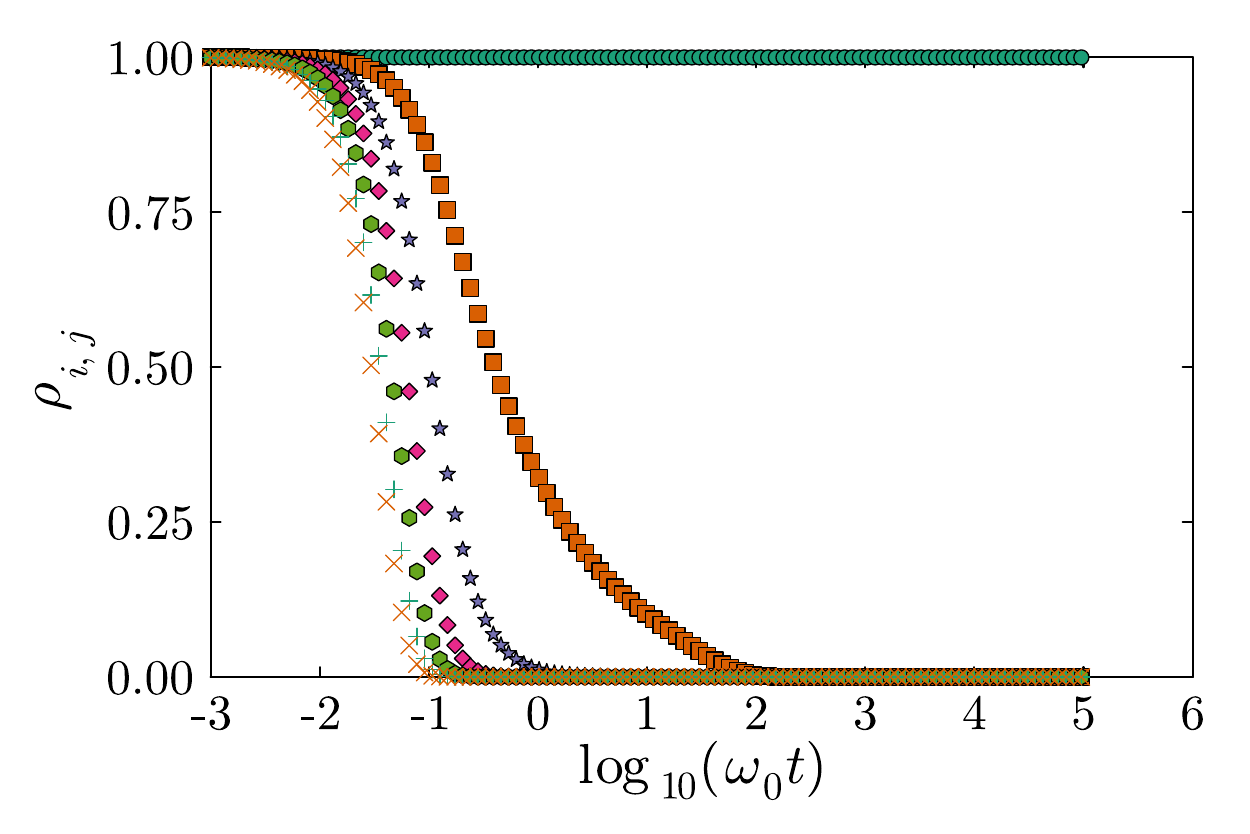}
         \caption{}
     \end{subfigure}
     \begin{subfigure}[b]{0.35\textwidth}
         \centering
          \stackinset{r}{0.001cm}{b}{0.01cm}{\includegraphics[width=0.2\textwidth]{label.png}}
         {\includegraphics[width=\textwidth]{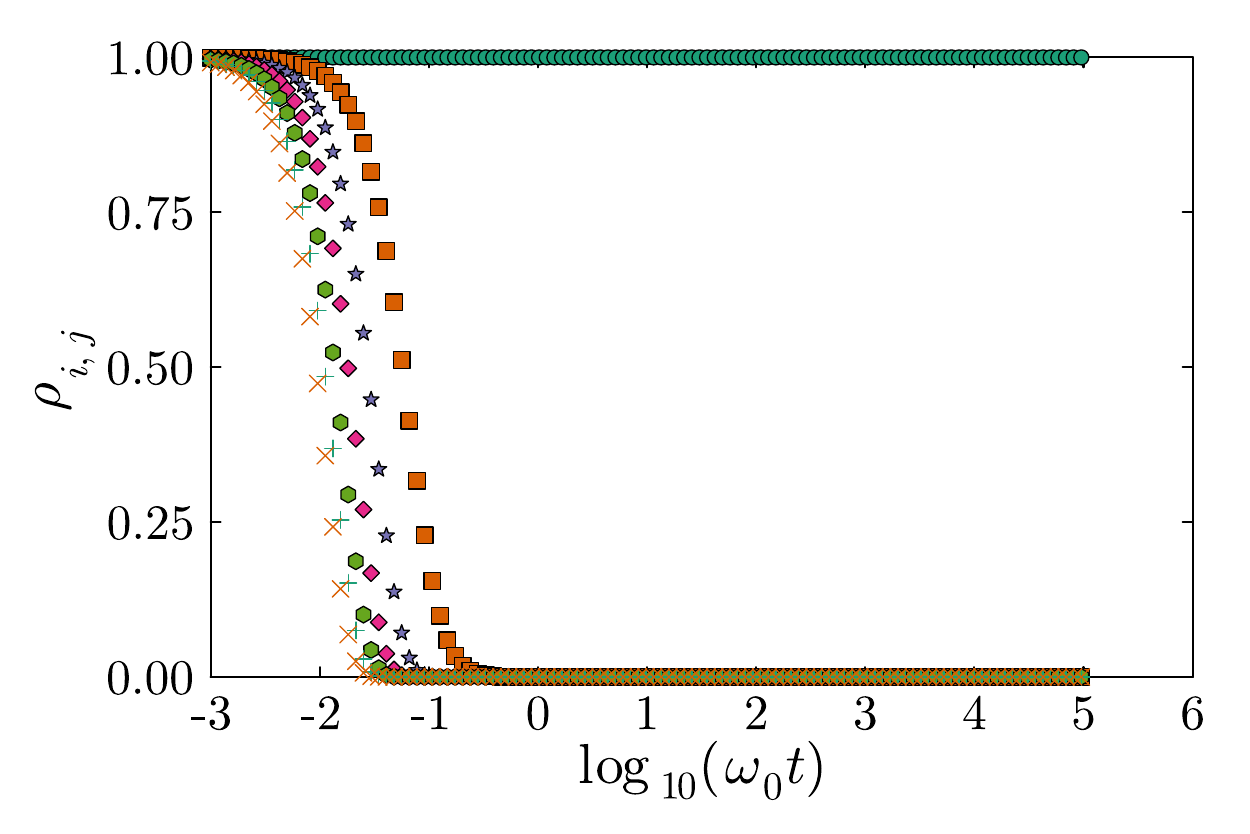}}
         \caption{}
     \end{subfigure}
\caption{Time decay of density matrix elements: $\rho_{1,1}(t)$ (circle), $\rho_{2,1}(t)$ (rectangle), $\rho_{3,1}(t)$ (five-point star), $\rho_{4,1}(t)$ (diamond), $\rho_{5,1}(t)$ (hexagon), $\rho_{6,1}(t)$ (cross), and $\rho_{7,1}(t)$ (x-cross). Subfigures show results for different temperatures: (a)~$\Gamma=0.01$, (b)~$\Gamma=0.1$, (c)~$\Gamma=1.0$, and (d)~$\Gamma=10.0$.}
\label{fig: gamma-elem}
\end{figure}
Another important parameter is $\Gamma$, which indicates how strongly the system interacts with the environment. We plot $P(t)$ and $D(t)$ in the top panel, and $C_e(\rho)$ and $C_2(\rho)$ in the bottom panel as functions of time for different values of $\Gamma$ (see Fig.~\ref{fig: gamma-norm}). We observe that decoherence starts to occur more rapidly while increasing $\Gamma$. For $\Gamma < 1$, as shown in Fig.~\ref{fig: gamma-norm}(a) and Fig.~\ref{fig: gamma-norm}(b), the decoherence time is still large ($\omega_0t \approx 10^2 - 10^4$). But for $\Gamma \ge 1$, as seen in Fig.~\ref{fig: gamma-norm}(c) and Fig.~\ref{fig: gamma-norm}(d), the decoherence time becomes very small ($\omega_0t \approx 1$). A higher $\Gamma$ means a stronger effect of the environment on the system, so decoherence happens faster. Also, for $P(t)$, we see that the oscillatory behaviour becomes weaker for larger $\Gamma$, especially when $\Gamma \ge 1$.

We also show the temporal behaviour of different density matrix elements in Fig.~\ref{fig: gamma-elem}(a)-(d) while increasing $\Gamma$, similar to what was done for temperature. The terms closer to the diagonal decay more slowly. As $\Gamma$ increases, the decay of all terms becomes faster. For $\Gamma=10$ in Fig.~\ref{fig: gamma-elem}(d), all terms decay before $\omega_0t=1.0$. For increasing $\Gamma$, the difference in decay times between the various terms ceases. \\
Decoherence without dissipation has already been studied for a harmonic oscillator \cite{gangopadhyay2001dissipationless}. By using the harmonic limit of the Morse potential, we observe that all quantities behave similarly to those for the Morse potential. The temporal behavior of $P(t)$, $D(t)$, $C_e(\rho)$, and $C_2(\rho)$ appears to be quite similar in both cases. The main difference, however, lies in the decoherence time, which is defined as $\tau_{element}$.\\
\begin{table}[h!]
\centering
\begin{tabular}{ccc}
    \toprule
    \multicolumn{3}{c}{$\Gamma = 0.01, \omega_c = 10$} \\
    \textbf{$kT$} & \textbf{Harmonic} & \textbf{Morse} \\
    \midrule
    0.01 & 27.06 & 526.715 \\
    0.10 & 6.818 & 68.79 \\
    1.00 & 1.68 & 9.02 \\
    \bottomrule
\end{tabular}
\hfill
\begin{tabular}{ccc}
    \toprule
    \multicolumn{3}{c}{$kT = 0.1 \omega_c = 10$} \\
    \textbf{$\Gamma$} & \textbf{Harmonic} & \textbf{Morse} \\
    \midrule
    0.01 & 6.818 & 68.79 \\
    0.10 & 0.077 & 0.43 \\
    1.00 & 0.02 & 0.04 \\
    \bottomrule
\end{tabular}
\caption{$\tau_{\text{element}}$ for different parameters under different limits}
\label{table: element}
\end{table}
We vary $kT$ and $\Gamma$ one at a time while keeping other parameters fixed. Significant changes are observed in both decoherence times, as shown in Table~\ref{table: element}. For $\tau_{element}$ in Table~\ref{table: element}, the decoherence time for the Morse oscillator (MO) is almost 10 times longer than that of the harmonic oscillator (HO). As seen in Fig.~\ref{fig: temp-norm}, Fig.~\ref{fig: temp-elem}, Fig.~\ref{fig: gamma-norm}, and Fig.~\ref{fig: gamma-elem}, increasing $kT$ and $\Gamma$ leads to faster decoherence, meaning a shorter decoherence time.
\section{Conclusion}
This study has investigated the phenomenon of dissipationless decoherence within the Morse oscillator (MO), a system offering a realistic model for molecular vibrations due to its inherent anharmonicity. Moving beyond traditional decoherence studies focused on energy dissipation, we explored how quantum coherence can be lost solely through system-environment interactions that preserve energy, a mechanism particularly relevant in low-dissipation molecular environments.

Our simulations, analysing measures such as survival probability $P(t)$, purity $D(t)$, relative entropy of coherence $C_e(\rho)$, and the 2-norm of coherence $C_2(\rho)$, confirm that dissipationless decoherence occurs in the MO, driven significantly by environmental temperature ($kT$) and coupling strength ($\Gamma$). Higher temperatures and stronger coupling were found to accelerate coherence loss, which aligns with our general expectation about decoherence processes. A key finding is the pronounced role of anharmonicity: the MO, with its non-equidistant energy levels converging towards a dissociation limit, exhibits markedly different decoherence dynamics compared to the harmonic oscillator (HO). Notably, the MO demonstrates significantly longer coherence times, with the element-wise coherence time ($\tau_{\text{element}}$) being approximately ten times longer than its harmonic counterpart under comparable conditions. This suggests an intrinsic protection against certain environmental noise afforded by the MO's structure.

Furthermore, we observed position-dependent decay rates for the off-diagonal elements of the density matrix ($\rho_{mn}(t)$), a feature more pronounced in the MO than the HO. Elements closer to the diagonal decay more slowly, and this hierarchy is influenced by both $kT$ and $\Gamma$. This nuanced decay behaviour underscores that different quantum transitions within the MO are affected differently by the environment, offering potential pathways for targeted quantum control.

The findings contribute to a deeper fundamental understanding of the quantum-to-classical transition, highlighting that it can be driven by the nature of system-environment entanglement even without energy exchange. For molecular physics, this work provides a more accurate framework for understanding coherence loss in spectroscopy and dynamics experiments, especially in regimes like low-pressure gases or molecular beams.

From a technological perspective, the enhanced coherence times of the MO suggest its potential utility in quantum information processing, potentially serving as a basis for more robust qubits or sensitive quantum sensors. Position-dependent decoherence rates could enable fine-grained control over specific molecular transitions, advancing quantum state engineering.

Future work could extend this model by incorporating more complex environmental spectral densities beyond the Ohmic approximation, investigating the interplay between dissipative and dissipationless mechanisms, exploring multi-dimensional potentials for polyatomic molecules, and developing experimental validation strategies using techniques like time-resolved spectroscopy on ultracold molecules.

In summary, this research illuminates the critical role of anharmonicity in dissipationless decoherence within the Morse oscillator. It advances our understanding of decoherence beyond purely dissipative models and highlights the potential of anharmonic molecular systems for quantum technologies, contingent on harnessing their unique coherence properties.
\section{Acknoweldegment}
TM acknowledges the financial support from UGC,
Government of India and IISER Kolkata for facilitating her research work. AA acknowledges the support and hospitality at VNIT Nagpur during the period of this research. SD is grateful to the Indian Institute of Science Education and Research, Kolkata, for inviting him to give a set of lectures, which provided a forum for initiating this project. He would also like to thank the Indian National Science Academy for an Honorary Scientist position that facilitated this collaboration. 
\bibliographystyle{apsrev4-1}
\bibliography{Morse}

\end{document}